\documentclass{article}

\usepackage{graphicx} 
\usepackage{threeparttable}
\usepackage[numbers,sort&compress,longnamesfirst,sectionbib]{natbib}
\usepackage{amsmath}
\usepackage{breqn}
\usepackage{epstopdf}
\usepackage[table]{xcolor}
\usepackage{algorithm}
\usepackage[noend]{algpseudocode}
\usepackage{tablefootnote}
\usepackage{subfigure}
\usepackage{lscape}
\usepackage{changepage}
\usepackage{multirow}
\usepackage{hhline}
\usepackage{hyperref}

\newcommand{\cellbluelight}{{\cellcolor{blue!15}}}

\newcommand{\ignore}[1]{}

\newcommand{\ad}[2]{\ensuremath{{\texttt{AllDifferent}}\!\left[#1,\ldots,#2\right]}}
\newcommand{\el}[2]{\ensuremath{{\texttt{Element}}\!\left(#1,#2\right)}}
\newcommand{\overbar}[1]{\mkern 1.5mu\overline{\mkern-1.5mu#1\mkern-1.5mu}\mkern 1.5mu}
\DeclareMathOperator*{\mycup}{\cup}

\begin{document}

\title{Exact Approaches for the Travelling Thief Problem}
\date{}

\author{
Junhua Wu \and Markus Wagner \and Sergey Polyakovskiy \and Frank Neumann \\
\\
Optimisation and Logistics\\
School of Computer Science \\
The University of Adelaide, Adelaide SA, Australia
}

\maketitle
\begin{abstract}
Many evolutionary and constructive heuristic approaches have been introduced in order to solve the Traveling Thief Problem (TTP). However, the accuracy of such approaches is unknown due to their inability to find global optima. In this paper, we propose three exact algorithms and a hybrid approach to the TTP. We compare these with state-of-the-art approaches to gather a comprehensive overview on the accuracy of heuristic methods for solving small TTP instances. 
\end{abstract}

\section{Introduction}
The travelling thief problem (TTP)~\cite{Bonyadi2013} is a recent academic problem in which two well-known combinatorial optimisation problems interact, namely the travelling salesperson problem (TSP) and the 0-1 knapsack problem (KP). It reflects the complexity in real-world applications that contain more than one $\mathcal{NP}$-hard problem, which can be commonly observed in the areas of planning, scheduling and routing. For example, delivery problems usually consist of a routing part for the vehicle(s) and a packing part of the goods onto the vehicle(s).

Thus far, many approximate approaches have been introduced for addressing the TTP and most of them are evolutionary or heuristic ~\cite{Polyakovskiy2017424}. Initially, \citet{Polyakovskiy2014TTP} proposed two iterative heuristics, namely the Random Local Search (RLS) and (1+1)-EA, based on a general approach that solves the problem in two steps, one for the TSP and one for the KP. Bonyadi et al.~\citep{Bonyadi2014gecco} introduced a similar two-phased algorithm named Density-based Heuristic (DH) and a method inspired by coevolution-based approaches named CoSolver. 
Mei et al.~\citep{Mei2014, Mei2015, Mei2016} also investigated the interdependency and proposed a cooperative coevolution based approach similar to CoSolver, and a memetic algorithm called MATLS that attempts to solve the problem as a whole. In 2015, Faulkner et al.~\cite{Faulkner2015} outperformed the existing approaches by their new operators and corresponding series of heuristics (named S1--S5 and C1--C6). Recently, \citet{Wagner2016} investigated the Max-Min Ant System (MMAS) \cite{Stutzle2000889} on the TTP, and \citet{ElYafrani:2016:PVS:2908812.2908847} proposed a memetic algorithm (MA2B) and a simulated annealing algorithm (CS2SA). The results show that the new algorithms were competitive to the state-of-the-art on a different range of TTP instances. Wagner et al.~\cite{Wagner2017} found in a study involving 21 approximate TTP algorithms that only a small subset of them is actually necessary to form a well-performing algorithm portfolio.

However, due to the lack of exact methods, all of the above-mentioned approximate approaches cannot be evaluated with respect to their accuracy even on small TTP instances. To address this issue, we propose three exact techniques and additional benchmark instances, which help to build a more comprehensive review of the approximate approaches. 

In the remainder, we revisit the definition of the TTP in Section~\ref{sec:ttp} and introduce our exact approaches in Section~\ref{sec:exact}. In Section~\ref{sec:experiments}, we elaborate on the setup of our experiments and compare our exact and hybrid approaches with the best approximate ones. The conclusions are drawn in Section~\ref{sec:conclusion}.

\section{Problem Statement}\label{sec:ttp}

In this section, we outline the problem formulation. For a comprehensive description, we refer the interested reader to~\cite{Polyakovskiy2014TTP}. 

Given is a set of cities $N=\{1,\ldots,n\}$ and a set of items $M=\{1,\ldots,m\}$. City $i$, $i=2,\ldots,n$, contains a set of items $M_i=\{1,\ldots,m_i\}$, \mbox{$M = \displaystyle \mycup_{i \in N} M_i$}. Item $k$ positioned in the city $i$ is characterised by its profit $p_{ik}$ and weight $w_{ik}$. The thief must visit each of the cities exactly once starting from the first city and return back to it in the end. The distance $d_{ij}$ between any pair of cities $i,j \in N$ is known. Any item may be selected as long as the total weight of collected items does not exceed the capacity $C$. A renting rate $R$ is to be paid per each time unit taken to complete the tour. $\upsilon_{max}$ and $\upsilon_{min}$ denote the maximal and minimum speeds of the thief. Assume that there is a binary variable $y_{ik} \in \left\{0,1\right\}$ such that $y_{ik} = 1$ iff item $k$ is chosen in city $i$. The goal is to find a tour $\Pi=\left(x_1,\ldots,x_n\right)$, $x_i \in N$, along with a packing plan $P= \left(y_{21},\ldots,y_{nm_n}\right)$ such that their combination $[\Pi,P]$ maximises the reward given in the form the following objective function. 
\begin{align}
Z([\Pi,P]) = & \displaystyle\sum_{i=1}^n \displaystyle\sum_{k=1}^{m_i} p_{ik} y_{ik} - R \left( \frac{d_{x_n x_1}}{\upsilon_{max}-\nu W_{x_n}} + \displaystyle\sum_{i = 1}^{n-1}\frac{d_{x_i x_{i+1}}}{\upsilon_{max}-\nu W_{x_i}} \right) \label{obj}
\end{align}
\noindent where $\nu = \left(\upsilon_{max}-\upsilon_{min}\right)/C$ is a constant value defined by input parameters. The minuend is the sum of all packed items' profits and the subtrahend is the amount that the thief pays for the knapsack's rent equal to the total traveling time along $\Pi$ multiplied by $R$. In fact, the actual travel speed along the distance $d_{x_i x_{i+1}}$ depends on the accumulated weight $W_{x_i}=\sum_{j=1}^i{\sum_{k=1}^{m_j} w_{jk} y_{jk}}$ of the items collected in the preceding cities $1,\ldots,i$. This then slows down the thief and has an impact on the overall benefit $Z$. 

\section{Exact Approaches to the TTP}\label{sec:exact}

In this section, we propose three exact approaches to the TTP. 

As a simplified version of the TTP, \citet{Polyakovskiy2017424} have recently introduced the packing while travelling problem (PWT), in which the tour $\Pi$ is predefined and only the packing plan $P$ is variable. Furthermore, Neumann et al.~\cite{2017arXiv170205217N} prove that the PWT can be solved in  pseudo-polynomial time by dynamic programming taking into account the fact that the weights are integer. The dynamic programming algorithm maps every possible weight $w$ to a packing plan $P$, i.e. $f:w \mapsto P$, which guarantees a certain profit. Then the optimal packing plan $P^*$ is to be selected among all the plans that have been obtained.  

Here, we adopt these findings to derive two exact algorithms for the TTP. Let $\cdot$ denote all possible weights for a given TTP instance. Let $[\Pi, f(\cdot)]$ designate the best solutions for the instance with tour $\Pi$ obtained via the dynamic programming for the PWT. As $\Pi$ is to be variable, the optimum objective value of the TTP is $Z^* = \arg \max_{\forall\Pi, w \in \cdot} Z([\Pi, f(w)]), $. This yields the basis for two of our approaches: dynamic programming (DP) and branch and bound search (BnB). The following sections describe the two approaches as well as a constraint programming (CP) technique adopted for the TTP.

\subsection{Dynamic Programming}\label{sec:dp2ttp}

Our DP approach is based on the Held-Karp algorithm for the TSP~\cite{Held:1961:DPA:800029.808532} and on the dynamic programming to the PWT~\cite{2017arXiv170205217N}. Algorithm~\ref{alg:dp} depicts the pseudocode for our approach. Let $\dot{S}=N\setminus\{1\}$ be a subset of the cities and $k\in N$ refer to a particular city. Then $[S,k]$ is a tour starting in city $1$, visiting all the cities in $S \subseteq \dot{S}$ exactly once, and ending in city $k$. The optimal solution of the TTP therefore can be described by $[\dot{S}, 1, f_{x_n}(W_{x_n})]^*$, where $W_{x_n}$ is the total weight of the knapsack when leaving the last city $x_n$ and $f_{x_n}$ results from the dynamic programming algorithm for the PWT considering the tour $\Pi=\left(1,\ldots,x_n\right)$. The following statement is valid with respect to the TTP's statement:
\begin{dmath*}
Z([\dot{S}, 1, f_{x_n}(W_{x_n})]^*) = Z([\dot{S}\setminus\{x_n\}, x_n, f_{x_{n-1}}(W_{x_n} - \overbar{W}_{x_n})]) +  \overbar{P}_{x_n}
	+ R \left ( \frac{d_{x_nx_1}}{v_{max} - \nu W_{x_n}} \right).
\end{dmath*}
Here, $\overbar{W}_{x_n}$ and $\overbar{P}_{x_n}$ are the total weight and the total profit of the items picked in city $x_n$. Clearly, $Z([\dot{S}\setminus\{x_n\}, x_n, f_{x_{n-1}}(W_{x_n} - \overbar{W}_{x_n})])$ is optimal for the tour \mbox{$[\dot{S}\setminus\{x_n\}, x_n]$}. Furthermore, such a relationship exists for every pair of $[S, i, f_j(W_{j})]$ and $[S\setminus\{j\}, j, f_{j-1}(W_{j} - \overbar{W}_j)]$, where $i \in \dot{S} \setminus S $ and $j \in S$. In fact, having an optimal solution for a given TTP instance, one can compute the optimal solution for the instance that excludes the last city from the solution of the original problem. Following this idea, we build our DP for the TTP.

The DP is costly in terms of the memory consumption, which reaches $\mathcal{O}(2^nnC)$. To reduce these cost, let $E_U$ define an upper bound on the value of a feasible solution built on the partial solution $[S, k, f_k(\cdot)]$ as follows:
$$
E_U([S, k, f_k(\cdot)]) = \max Z([S, k, f_{k}(\cdot)]) + \sum\limits_{i \in \dot{S} \setminus S} \sum\limits_{j=1}^{m_i} p_{ij} - R\frac{d_{k1}}{v_{max}}
$$
It estimates the maximal profit that the thief may obtain by passing the remaining part of the tour with the maximal speed, that is generating the minimal possible cost. Obviously, this guarantees that the complete optimal solution can not exceed the bound. Therefore, if any incumbent solution is known, it is valid to eliminate a partial solution $[S, k, f_k(w)], w \in \cdot$ if $E_U([S, k, f_k(w)]) < Z_L$, where $Z_L$ is the objective value of the incumbent. In practice, one can obtain an incumbent solution (and compute $Z_L$) in two stages. First, a feasible solution $\Pi$ for the TSP part of the problem can be computed by solvers such as Concorde~\cite{cook2005concorde} or by the Lin-Kernighan algorithm~\cite{lin1973effective}. Second, the dynamic programming applied for $\Pi$ contributes the packing plan.

\begin{algorithm}[!t]
\caption{Dynamic programming to the TTP}\label{alg:dp}
\begin{algorithmic}[1]
	\Procedure{Dynamic programming}{}
		\State $\dot{S} = \{2,...,n\}$, $S=\emptyset, w = 0$
		\For{ $i = 2$ to $n$ }
			\State store $[S, i, w] = Z([S, i, f_i(w)])$ in a mapping $M$
		\EndFor
		\For{ $i = 3$ to $n$ }
			\State $M' = M$
			\For{all $[S, k, w]$ in $M'$}
				\For {all $\dot{s} \in \dot{S} \setminus S$}
					\For {all $s \in S$}
						\State calculate  $Z([S \cup \{\dot{s}\}, \dot{s}, f_{\dot{s}}(\cdot)])$ from $Z([S, s, f_{s}(\cdot)])$
						\State and store $[S \cup \{\dot{s}\}, \dot{s}, \cdot] = \max Z([S \cup \{\dot{s}\}, \dot{s}, f_{\dot{s}}(\cdot)])$ in $M'$
					\EndFor
				\EndFor
			\EndFor
			\State $M = M'$
		\EndFor
		\State calculate $Z([\dot{S}, 1, f_1(W_{x_n})]^*)$ from $Z([\dot{S} \setminus \{i\}, i, f_{i}(\cdot)]^*), i \in \{2,...,n\}$ in $M$
	\EndProcedure
\end{algorithmic}
\end{algorithm}

\subsection{Branch and Bound Search}\label{sec:bnb}

Now, we introduce a branch and bound search for the TTP employing the upper bound $E_U$ defined in Section~\ref{sec:dp2ttp}. Algorithm~\ref{alg:bnb} depicts the pseudocode, where $\Pi_i, i \in \{1,...,n\}$ denotes a sub-permutation of $\Pi$ with the cities $1$ to $i$ visited, and $f_{i}$ is the mapping $f:w \mapsto P$ calculated for $\Pi$ by the dynamic programming for the PWT.

A way to tighten the upper bound $E_U$ is by providing a better estimation of the remaining distance from the current city $k$ to the last city of the tour. Currently, the shortest distance from $k$ to $1$, i.e. $d_{k1}$, is used. The following two ways can improve the estimation: (i) the use of distance $d_{f1}$ from city $f$ to city $1$, where $f$ is the farthest unvisited city from $1$; (ii) the use of distance $d^* - d_t$, where $d^*$ is the shortest path that can be pre-calculated and $d_t$ is the distance passed so far to achieve city $k$ in the tour $\Pi$. These two ideas can be joined together by using the $\max\{d_f,(d^*-d_t)\}$ to enhance the result.

\begin{algorithm}[!t]
\caption{Branch and Bound Search for the TTP}\label{alg:bnb}
\begin{algorithmic}[1]
	\Procedure{BnB Search}{}
		\State Create an initial solution to gain the benefit $best$ and an tour permutation $\Pi$
		\State Create an empty mapping $M$
		\State Set $l=0$
		\State \Call{Search}{$\Pi, l, M, best$}
	\EndProcedure
	\Function{Search}{$\Pi, l, M, best$}
		\If{$l == n$}
			\State calculate $ Z([\Pi, f_n(\cdot)])$ from $Z([\Pi_{n-1}, f_{n-1}(\cdot)])$ in $M$
			\State \Return $\max\{\max{Z([\Pi, f_n(\cdot)])}, best\}$ 
		\Else
		
		\For{ $i = l+1$ to $n$ }
			\State Swap cities $l+1$ and $i$ in $\Pi$
			\State Set M' = Calculate $ Z([\Pi_{l+1}, f_{l+1}(\cdot)])$ from $Z([\Pi_{l}, f_{l}(\cdot)])$ in $M$
			\If{$\max E_U([\Pi_{l+1}, f_{l+1}(\cdot)]) > best$}
			\State $best = \max\{ best,$ \Call{Search}{$\Pi, l+1, M', best$} $\}$
			\EndIf
			\State Swap cities $l+1$ and $i$ in $\Pi$
		\EndFor
		\State \Return $best$
		\EndIf
		
	\EndFunction
\end{algorithmic}
\end{algorithm}

\subsection{Constraint Programming}\label{sec:constrant}

Now, we present our third exact approach adopting the existing state-of-the-art constraint programming (CP) paradigm \citep{H02}. Our model employs a simple permutation based representation of the tour that allows the use of the \texttt{AllDifferent} filtering algorithm \citep{Benchimol2012}. Similarly to the Section~\ref{sec:ttp}, a vector $W = \left(W_1,\ldots,W_n\right)$ is used to refer to the total weights accumulated in the cities of tour $\Pi$. Specifically, $W_i$ is the weight of the knapsack when the thief departs from city $i$. The model bases the search on two types of decision variables:
\begin{itemize}
\item $\textbf{\textit{x}}$ denotes the particular positions of the cities in tour $\Pi$. Variable $x_i$ takes the value of $j \in N$ to indicate that $j$ is the $i$th city to be visited. The initial variable domain of $x_1$ is $D\left(x_1\right)=\left\{1\right\}$ and it is $D\left(x_i\right)=N\setminus \left\{i\right\}$ for any subsequently visited city $i=2,\ldots,n$.
\item $\textbf{\textit{y}}$ signals on the selection of an item in the packing plan $P$. Variable $y_{ik}$, $i \in N$, $k \in M_i$, is binary, therefore $D\left(y_{ik}\right)=\left\{0,1\right\}$.
\end{itemize}
Furthermore, an integer-valued vector $d$ is used to express the distance matrix so that its element $n\left(x_i-1\right) + x_{i+1}$ equals the distance $d_{x_ix_{i+1}}$ between two consecutive cities $x_i$ and $x_{i+1}$ in $\Pi$. The model relies on the $\ad{x_1}{x_n}$ constraint, which ensures that the values of $x_1,\ldots,x_n$ are distinct. It also involves the $\el{g}{h}$ expression, which returns the $h$th variable in the list of variables $g$. In total, the model (CPTTP) consists of the following objective function and constraints:

{\footnotesize
\begin{flalign}
\nonumber max\, 
& \sum\limits_{i=1}^n \sum_{j=1}^{m_i} p_{ij}y_{ij}
\\
&
- R \!\left( \sum_{i=1}^{n-1} \frac{\el{d}{n\left(x_i-1\right) + x_{i+1}}}{v_{max} - \nu \el{W}{x_i}} + \frac{\el{d}{n \left(x_n-1\right)+1}}{v_{max} - \nu \el{W}{x_n}}\right)
\label{cp:obj}
\\
&
\ad{x_1}{x_n}
\label{cp:1}
\\
&
W_i = W_{i-1} + \sum_{j \in M_i} w_{ij} y_{ij}, \,\, i \in \left\{2,\ldots,n\right\}
\label{cp:2}
\\
&
W_n \leq C
\label{cp:3}
\end{flalign}
}

Expression~(\ref{cp:obj}) calculates the objective value according to function~(\ref{obj}). Constraint~(\ref{cp:1}) verifies that all the cities are assigned to different positions, and thus are visited exactly once. This is a sub-tour elimination constraint. Equation (\ref{cp:2}) calculates the weight $W_i$ of all the items collected in the cities $1,\ldots,i$. Equation~(\ref{cp:3}) is a capacity constraint. 

The performance of a CP model depends on its solver; specifically, on the filtering algorithms and on the search strategies it applies. Here, we use IBM ILOG \textsc{CP Optimizer} 12.6.2 with its searching algorithm set to the \textit{restart mode}. This mode adopts a general purpose search strategy~\cite{Refalo2004} inspired from integer programming techniques and is based on the concept of the impact of a variable. The impact measures the importance of a variable in reducing the search space. The impacts, which are learned from the observation of the domains' reduction during the search, help the restart mode dramatically improve the performance of the search. Within the search, the cities are assigned to the positions first and then the items are decided on. Therefore, the solver instantiates $x_1,\ldots,x_n$ prior to $y_{21},\ldots,y_{nm_n}$ variables applying its default selection strategy. Our extensive study shows that such an order gives the best results fast.

\section{Computational Experiments}\label{sec:experiments}

In this section, we first compare the performance of the exact approaches to TTP in order to find the best one for setting the baseline for the subsequent comparison of the approximate approaches. 
Our experiments run on the CPU cluster of the Phoenix HPC at the University of Adelaide, which contains 3072 Intel(R) Xeon(R) 2.30GHz CPU cores and 12TB of memory. We allocate one CPU core and 32GB of memory to each individual experiment.

\subsection{Computational Set Up}\label{sec:benchmark}

\begin{table}[ht]
\centering
\renewcommand{\arraystretch}{1.05}
\setlength{\tabcolsep}{2.9mm}
\begin{tabular}{|l|ll|l|l|l|}
\hline
&&&  \multicolumn{3}{c|}{Running time (in seconds)} \\
\cline{4-6}
Instance & n	&	m	&	DP	&	BnB & CP \\
 \hline
eil51\_n05\_m4\_uncorr\_01	&	5	&	4	&	0.018	&	0.023	&	0.222	\\
eil51\_n06\_m5\_uncorr\_01	&	6	&	5	&	0.07	&	0.079	&	0.24	\\
eil51\_n07\_m6\_uncorr\_01	&	7	&	6	&	0.143	&	0.195	&	0.497	\\
eil51\_n08\_m7\_uncorr\_01	&	8	&	7	&	0.343	&	0.505	&	4.594	\\
eil51\_n09\_m8\_uncorr\_01	&	9	&	8	&	0.633	&	1.492	&	63.838	\\
eil51\_n10\_m9\_uncorr\_01	&	10	&	9	&	0.933	&	5.188	&	776.55	\\
eil51\_n11\_m10\_uncorr\_01	&	11	&	10	&	2.414	&	23.106	&	12861.181	\\
eil51\_n12\_m11\_uncorr\_01	&	12	&	11	&	3.938	&	204.786	&	-	\\
eil51\_n13\_m12\_uncorr\_01	&	13	&	12	&	14.217	&	2007.074	&	-	\\
eil51\_n14\_m13\_uncorr\_01	&	14	&	13	&	13.408	&	36944.146	&	-	\\
eil51\_n15\_m14\_uncorr\_01	&	15	&	14	&	89.461	&	-	&	-	\\
eil51\_n16\_m15\_uncorr\_01	&	16	&	15	&	59.526	&	-	&	-	\\
eil51\_n17\_m16\_uncorr\_01	&	17	&	16	&	134.905	&	-	&	-	\\
eil51\_n18\_m17\_uncorr\_01	&	18	&	17	&	366.082	&	-	&	-	\\
eil51\_n19\_m18\_uncorr\_01	&	19	&	18	&	830.18	&	-	&	-	\\
eil51\_n20\_m19\_uncorr\_01	&	20	&	19	&	2456.873	&	-	&	-	\\
\hline
\end{tabular}
\caption{Columns `n' and `m' denote the number of cities and the number of items, respectively. Running times are given in seconds for DP, BnB and CP for different numbers of cities and items. `-' denotes the case when an approach failed to achieve an optimal solution in the given time limit.}
\label{tab:runningtime}
\end{table}

To run our experiments, we generate an additional set of small-sized instances following the way proposed in~\cite{Polyakovskiy2014TTP}\footnote{All instances are available online: \url{http://cs.adelaide.edu.au/~optlog/research/ttp.php}}. We use only a single instance of the original TSP library~\cite{tsplib} as the starting point for our new subset. It is entitled as \texttt{eil51} and contains 51 cities. Out of these cities, we select uniformly at random cities that we removed in order to obtain smaller test problems with $n=5,\dots,20$ cities. To set up the knapsack component of the problem, we adopt the approach given in~\cite{Pisinger2005} and use the corresponding problem generator available in~\cite{Pisinger20052271}. As one of the input parameters, the generator asks for the range of coefficients, which we set to 1000. In total, we create knapsack test problems containing $k(n-1)$, $k\in\left\{1,5,10\right\}$ items and which are characterised by a knapsack capacity category $Q \in \left\{1,6,10\right\}$. Our experiments focus on \textit{uncorrelated} (\texttt{uncorr}), \textit{uncorrelated with similar weights} (\texttt{uncorr-s-w}), and \textit{multiple strongly correlated} (\texttt{m-s-corr}) types of instances. At the stage of assigning the items of a knapsack instance to the particular cities of a given TSP tour, we sort the items in descending order of their profits and the second city obtains $k$, $k\in\left\{1,5,10\right\}$, items of the largest profits, the third city then has the next $k$ items, and so on. All the instances use the ``CEIL\_2D'' for intra-city distances, which means that the Euclidean distances are rounded up to the nearest integer. We set $\upsilon_{min}$ and $\upsilon_{max}$ to $0.1$ and $1$. 

Tables~\ref{tab:runningtime} and~\ref{tab:results2} illustrate the results of the experiments. The test instances' names should be read as follows. First, \texttt{eil51} stays for the name of the original TSP problem. The values succeeding $n$ and $m$ denote the actual number of cities and the total number of items, respectively, which are further followed by the generation type of a knapsack problem. Finally, the postfixes 1, 6 and, 10 in the instances' names describe the knapsack's capacity $C$.

\subsection{Comparison of the exact approaches}\label{sec:cdb}

We compare the three exact algorithms by allocating each instance a generous 24-hour time limit. Our aim is to analyse the running time of the approaches influenced by the increasing number of cities. Table~\ref{tab:runningtime} shows the running time of the approaches.

\subsection{Comparison between DP and Approximate Approaches}\label{sec:cda}
With the exact approaches being introduced, approximate approaches can be evaluated with respect to their accuracy to the optima. In the case of the TTP, most state-of-the-art approximate approaches are evolutionary algorithms and local searches, such as Memetic Algorithm with 2-OPT and Bit-flip (MA2B), CoSolver-based with 2-OPT, and Simulated Annealing (CS2SA) in \cite{ElYafrani:2016:PVS:2908812.2908847}, CoSolver-based with 2-OPT and Bit-flip (CS2B) in \cite{el2015cosolver2b}, and S1, S5, and C5 in \cite{Faulkner2015}. 

\subsubsection{Hybrid Approaches.}\label{sec:hybrid}
In addition to existing heuristics, we introduce enhanced approaches of S1 and S5, which are hybrids of the two and that one of dynamic programming for the PWT~\cite{2017arXiv170205217N}. The original S1 and S5 work as follows. First, a single TSP tour is computed using the Chained Lin-Kernighan-Heuristic~\cite{lin1973effective}, then a fast packing heuristic is applied. S1 performs these two steps only once and only in this order, while S5 repeats S1 until the time budget is exhausted. Our hybrids DP-S1 and DP-S5 are equivalent to these two algorithms, however, they use the exact dynamic programming to the PWT as a packing solver. This provides better results as we can now compute the optimal packing for the sampled TSP tours.

\subsubsection{Results.}

\begin{figure}
\centering
\includegraphics[width=130mm]{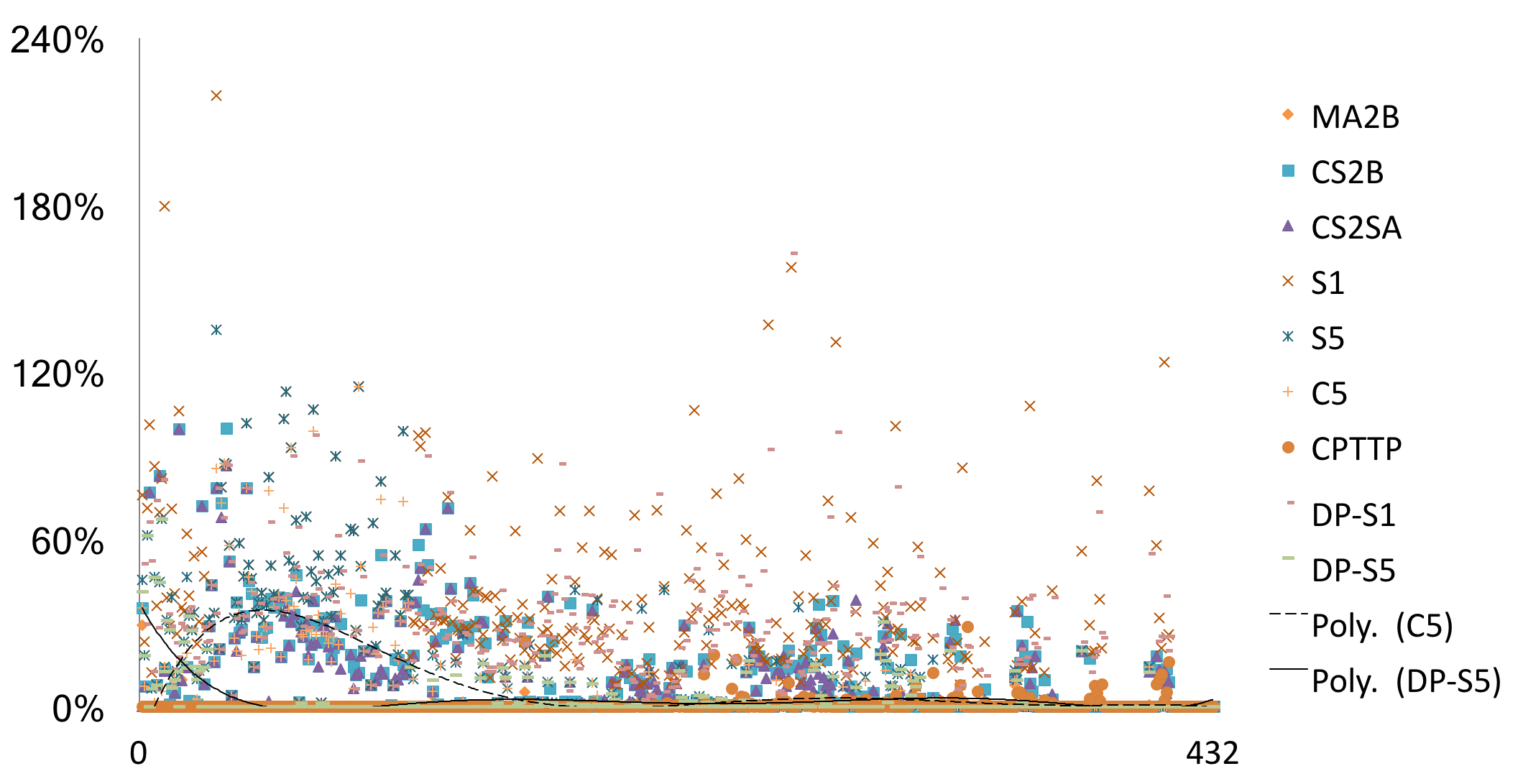}
\caption{Showing a gap to an optimal solution when one has been obtained by an exact approach. From left to right: the 432 instances first sorted by the number of cities, then by the total number of items.}
\label{fig:resultsCities}
\end{figure}

We start by showing a performance summary of 10 algorithms on 432 instances in Table~\ref{tab:broadOverview}. 
In addition, Table~\ref{tab:results2} shows detailed results for a subset of the best approaches on a subset of instances. 
Figure~\ref{fig:resultsCities} shows the results of the entire comparison. We include trend lines\footnote{They are fitted polynomials of degree six used only for visualisation purposes.} for two selected approaches, which we will explain in the following.

We would like to highlight the following observations:

\begin{enumerate}
\item S1 performs badly across a wide range of instances. Its restart variant S5 performs better, however, its lack of a local search becomes apart in its relatively bad performance (compared to other approaches) on small instances.
\item C5 performs better than both S1 and S5, which is most likely due to its local searches that differentiate it from S1 and S5. Still, we can see a ``hump'' in its trend line for smaller instances, which flattens out quickly for larger instances.
\item The dynamic programming variants DP-S1 and DP-S5 perform slightly better than S1 and S5, which shows the difference in quality of the packing strategy; however, this is at times balanced out by the faster packing which allows more TSP tours to be sampled. For small instances, DP-S5 lacks a local search on the tours, which is why its gap to the optimum is relatively large, as shown by the respective trend lines.
\item MA2B dominates the field with outstanding performance across all instances, independent of number of cities and number of items. Remarkable is the high reliability with which it reaches a global optimum.
\end{enumerate}

Interestingly, all approaches seem to have difficulties solving instances with the knapsack configuration multiple-strongly-corr\_01 (see Table~\ref{tab:results2}). Compared to the other two knapsack types, TTP-DP takes the longest to solve the strongly correlated ones. Also, these tend to be the only instances for which the heuristics rarely find optimal solutions, if at all.

\begin{table}[!t]%
\centering\small{\scalebox{1.0}{%
\renewcommand{\arraystretch}{1.06}%
\setlength{\tabcolsep}{2.3mm}%
\begin{tabular}{|lcccccccc|}\hline
gap \ignore{  & CPTTP  }      & MA2B        & CS2B        & CS2SA       & S1           & S5           & C5           & DP-S1       & DP-S5 \\ \hline 
avg                                \ignore{  & 1.0\% }  & 0.3\% & 15.3\% & 11.5\% & 38.9\%  & 15.7\%  & 09.9\%  & 30.1\% & 3.3\%  \\
stdev                              \ignore{  & 3.1\% } & 2.2\% & 17.8\% & 16.7\% & 29.4\%  & 24.6\%  & 18.8\%  & 20.1\% & 8.5\%  \\
\#$_{\text{opt}}$                  \ignore{   & 222  }        & 312         & 70          & 117         & 3            & 42           & 193          & 5           & 85           \\
\#$_{\text{1\%}}$ & 265            \ignore{  & 320   }      & 100         & 132         & 10           & 160          & 193          & 9           & 245          \\
\#$_{\text{10\%}}$ & 324            \ignore{ & 328  }       & 161         & 206         & 27           & 203          & 240          & 33          & 288          \\ \hline
\end{tabular}
}}
\caption{Performance summary of heuristic TTP solvers across all instances for which the optimal result has been obtained. \#$_{\text{opt}}$ is the number of times when the average of 10 independent repetitions is equal to the optimum. \#$_{\text{1\%}}$ and \#$_{\text{10\%}}$ show the number of times the averages are within 1\% and 10\%.}
\label{tab:broadOverview}
\end{table}

\begin{table}[!htbp]%
\centering%
\renewcommand{\arraystretch}{1.06}
\setlength{\tabcolsep}{.4mm}
\scriptsize
\begin{adjustwidth}{0cm}{}
\begin{tabular}{l|rr|rrr|rr|rr}
\hline
\multicolumn{1}{l|}{}                          & \multicolumn{2}{c|}{TTP-DP}                                        & \multicolumn{3}{c|}{MA2B}                                                                 & \multicolumn{2}{c|}{C5}                             & \multicolumn{2}{c}{DP-S5}                          \\ 
\multicolumn{1}{l|}{Instance}                 & \multicolumn{1}{r}{OPT} & \multicolumn{1}{r|}{RT} & \multicolumn{1}{r}{Gap} & \multicolumn{1}{r}{Std} & \multicolumn{1}{r|}{RT} & \multicolumn{1}{r}{Gap} & \multicolumn{1}{r|}{Std} & \multicolumn{1}{r}{Gap} & \multicolumn{1}{r}{Std} \\ 
\hline
eil51\_n05\_m4\_multiple-strongly-corr\_01	&	619.227	&	0.02	&	\cellbluelight29.1	&	12.1	&	2.71	&	35.5	&	1.20e-6	&	41.3	&	0.0	\\
eil51\_n05\_m4\_uncorr\_01	&	466.929	&	0.02	&	\cellbluelight0.0	&	0.0	&	3.22	&	\cellbluelight0.0	&	2.20e-6	&	\cellbluelight0.0	&	2.20e-6	\\
eil51\_n05\_m4\_uncorr-similar-weights\_01	&	299.281	&	0.02	&	\cellbluelight0.0	&	0.0	&	3.21	&	7.8	&	2.40e-6	&	7.8	&	1.20e-6	\\
eil51\_n05\_m20\_multiple-strongly-corr\_01	&	773.573	&	0.08	&	13.4	&	0.0	&	1.44	&	14.3	&	0.0	&	\cellbluelight12.8	&	0.0	\\
eil51\_n05\_m20\_uncorr\_01	&	2144.796	&	0.07	&	\cellbluelight0.0	&	0.0	&	3.35	&	7.4	&	0.0	&	6.6	&	2.30e-6	\\
eil51\_n05\_m20\_uncorr-similar-weights\_01	&	269.015	&	0.04	&	\cellbluelight0.0	&	0.0	&	3.51	&	\cellbluelight0.0	&	2.30e-6	&	\cellbluelight0.0	&	0.0	\\
eil51\_n10\_m9\_multiple-strongly-corr\_01	&	573.897	&	1.21	&	\cellbluelight0.0	&	0.0	&	6.07	&	\cellbluelight0.0	&	0.0	&	\cellbluelight0.0	&	0.0	\\
eil51\_n10\_m9\_uncorr\_01	&	1125.715	&	0.93	&	\cellbluelight0.0	&	0.0	&	6.06	&	\cellbluelight0.0	&	1.30e-6	&	\cellbluelight0.0	&	1.30e-6	\\
eil51\_n10\_m9\_uncorr-similar-weights\_01	&	753.230	&	0.86	&	\cellbluelight0.0	&	0.0	&	5.87	&	\cellbluelight0.0	&	0.0	&	\cellbluelight0.0	&	0.0	\\
eil51\_n10\_m45\_multiple-strongly-corr\_01	&	1091.127	&	14.89	&	\cellbluelight0.0	&	0.0	&	7.99	&	\cellbluelight0.0	&	0.0	&	\cellbluelight0.0	&	0.0	\\
eil51\_n10\_m45\_uncorr\_01	&	6009.431	&	6.39	&	\cellbluelight0.0	&	0.0	&	8.6	&	6.6	&	2.30e-6	&	\cellbluelight0.0	&	0.0	\\
eil51\_n10\_m45\_uncorr-similar-weights\_01	&	3009.553	&	8.87	&	\cellbluelight0.0	&	0.0	&	6.78	&	\cellbluelight0.0	&	2.30e-6	&	\cellbluelight0.0	&	2.30e-6	\\
eil51\_n12\_m11\_multiple-strongly-corr\_01	&	648.546	&	4.58	&	\cellbluelight0.0	&	0.0	&	6.08	&	4.6	&	2.20e-6	&	4.6	&	2.20e-6	\\
eil51\_n12\_m11\_uncorr\_01	&	1717.699	&	3.94	&	\cellbluelight0.0	&	0.0	&	7.21	&	\cellbluelight0.0	&	1.20e-6	&	\cellbluelight0.0	&	1.20e-6	\\
eil51\_n12\_m11\_uncorr-similar-weights\_01	&	774.107	&	3.36	&	\cellbluelight0.0	&	0.0	&	7.03	&	\cellbluelight0.0	&	2.30e-6	&	\cellbluelight0.0	&	2.30e-6	\\
eil51\_n12\_m55\_multiple-strongly-corr\_01	&	1251.780	&	117.99	&	\cellbluelight0.0	&	0.0	&	9.19	&	\cellbluelight0.0	&	0.0	&	\cellbluelight0.0	&	0.0	\\
eil51\_n12\_m55\_uncorr\_01	&	8838.012	&	35.79	&	\cellbluelight0.0	&	0.0	&	9.76	&	\cellbluelight0.0	&	0.0	&	\cellbluelight0.0	&	0.0	\\
eil51\_n12\_m55\_uncorr-similar-weights\_01	&	3734.895	&	38.36	&	12.3	&	0.0	&	8.34	&	12.3	&	0.0	&	\cellbluelight0.2	&	0.0	\\
eil51\_n15\_m14\_multiple-strongly-corr\_01	&	547.419	&	39.82	&	\cellbluelight0.0	&	0.0	&	7.87	&	14.1	&	1.30e-6	&	13.3	&	1.30e-6	\\
eil51\_n15\_m14\_uncorr\_01	&	2392.996	&	89.46	&	\cellbluelight0.0	&	0.0	&	7.28	&	3.8	&	0.0	&	3.8	&	0.0	\\
eil51\_n15\_m14\_uncorr-similar-weights\_01	&	637.419	&	16.35	&	\cellbluelight0.0	&	0.0	&	6.86	&	\cellbluelight0.0	&	1.60e-6	&	\cellbluelight0.0	&	1.60e-6	\\
eil51\_n15\_m70\_multiple-strongly-corr\_01	&	920.372	&	3984.29	&	2.1	&	1.1	&	12.11	&	\cellbluelight0.0	&	2.70e-6	&	\cellbluelight0.0	&	2.70e-6	\\
eil51\_n15\_m70\_uncorr\_01	&	9922.137	&	740.22	&	\cellbluelight0.0	&	0.0	&	9.67	&	7	&	1.20e-6	&	1.9	&	0.0	\\
eil51\_n15\_m70\_uncorr-similar-weights\_01	&	4659.623	&	867.78	&	\cellbluelight0.0	&	0.0	&	7.98	&	\cellbluelight0.0	&	0.0	&	\cellbluelight0.0	&	0.0	\\
eil51\_n16\_m15\_multiple-strongly-corr\_01	&	794.745	&	105.5	&	\cellbluelight0.0	&	0.0	&	7.7	&	18.9	&	1.6e-6	&	18.9	&	1.6e-6	\\
eil51\_n16\_m15\_multiple-strongly-corr\_10	&	4498.848	&	623.4	&	\cellbluelight0.0	&	0.0	&	9.1	&	12.9	&	0.0	&	16.6	&	1.3e-6	\\
eil51\_n16\_m15\_uncorr\_01	&	2490.889	&	59.5	&	\cellbluelight1.0	&	0.7	&	8.4	&	1.6	&	2.3e-6	&	1.6	&	2.3e-6	\\
eil51\_n16\_m15\_uncorr\_10	&	3601.077	&	211.5	&	\cellbluelight0.0	&	0.0	&	9.0	&	7.1	&	1.6e-6	&	7.1	&	1.6e-6	\\
eil51\_n16\_m15\_uncorr-similar-weights\_01	&	540.897	&	36.4	&	\cellbluelight0.0	&	0.0	&	8.5	&	\cellbluelight0.0	&	3.0e-6	&	\cellbluelight0.0	&	3.0e-6	\\
eil51\_n16\_m15\_uncorr-similar-weights\_10	&	3948.211	&	245.4	&	\cellbluelight0.0	&	0.0	&	8.7	&	5.8	&	1.5e-6	&	13.6	&	0.0	\\
eil51\_n17\_m16\_multiple-strongly-corr\_01	&	685.565	&	248.6	&	\cellbluelight0.0	&	0.0	&	8.4	&	0.2	&	1.5e-6	&	\cellbluelight0.0	&	1.5e-6	\\
eil51\_n17\_m16\_multiple-strongly-corr\_10	&	3826.098	&	2190.4	&	\cellbluelight0.0	&	0.0	&	9.8	&	\cellbluelight0.0	&	1.5e-6	&	\cellbluelight0.0	&	1.5e-6	\\
eil51\_n17\_m16\_uncorr\_01	&	2342.664	&	134.9	&	\cellbluelight0.0	&	0.0	&	8.3	&	\cellbluelight0.0	&	0.0	&	\cellbluelight0.0	&	0.0	\\
eil51\_n17\_m16\_uncorr\_10	&	2275.279	&	554.5	&	\cellbluelight0.0	&	0.0	&	9.6	&	\cellbluelight0.0	&	0.0	&	\cellbluelight0.0	&	0.0	\\
eil51\_n17\_m16\_uncorr-similar-weights\_01	&	556.851	&	70.8	&	\cellbluelight0.0	&	0.0	&	8.1	&	\cellbluelight0.0	&	0.0	&	\cellbluelight0.0	&	0.0	\\
eil51\_n17\_m16\_uncorr-similar-weights\_10	&	2935.961	&	787.7	&	\cellbluelight0.0	&	0.0	&	9.7	&	\cellbluelight0.0	&	0.0	&	\cellbluelight0.0	&	0.0	\\
eil51\_n18\_m17\_multiple-strongly-corr\_01	&	834.031	&	715.7	&	\cellbluelight7.9	&	0.8	&	10.2	&	9.2	&	0.0	&	12.9	&	1.7e-6	\\
eil51\_n18\_m17\_multiple-strongly-corr\_10	&	5531.373	&	6252.4	&	\cellbluelight0.0	&	0.0	&	10.5	&	0.4	&	1.5e-6	&	0.4	&	1.5e-6	\\
eil51\_n18\_m17\_uncorr\_01	&	2644.491	&	366.1	&	\cellbluelight0.0	&	0.0	&	9.7	&	0.2	&	0.0	&	1.8	&	0.0	\\
eil51\_n18\_m17\_uncorr\_10	&	3222.603	&	1462.7	&	\cellbluelight0.0	&	0.0	&	10.3	&	\cellbluelight0.0	&	1.3e-6	&	0.2	&	0.0	\\
eil51\_n18\_m17\_uncorr-similar-weights\_01	&	532.906	&	148.3	&	\cellbluelight0.0	&	0.0	&	8.5	&	\cellbluelight0.0	&	1.3e-6	&	\cellbluelight0.0	&	1.3e-6	\\
eil51\_n18\_m17\_uncorr-similar-weights\_10	&	4420.438	&	1929.3	&	\cellbluelight0.0	&	0.0	&	9.9	&	\cellbluelight0.0	&	2.9e-6	&	0.3	&	1.8e-6	\\
eil51\_n19\_m18\_multiple-strongly-corr\_01	&	910.229	&	1771.6	&	\cellbluelight0.0	&	0.0	&	9.3	&	20.1	&	1.6e-6	&	20.1	&	1.6e-6	\\
eil51\_n19\_m18\_multiple-strongly-corr\_10	&	-	&	-	&	-	&	-	&	10.4	&	-	&	-	&	-	&	-	\\
eil51\_n19\_m18\_uncorr\_01	&	2604.844	&	830.2	&	\cellbluelight0.0	&	0.0	&	9.7	&	\cellbluelight0.0	&	0.0	&	\cellbluelight0.0	&	0.0	\\
eil51\_n19\_m18\_uncorr\_10	&	4048.408	&	3884.3	&	\cellbluelight0.0	&	0.0	&	10.9	&	\cellbluelight0.0	&	1.4e-6	&	\cellbluelight0.0	&	1.4e-6	\\
eil51\_n19\_m18\_uncorr-similar-weights\_01	&	472.186	&	412.3	&	\cellbluelight0.0	&	0.0	&	9.2	&	\cellbluelight0.0	&	1.5e-6	&	\cellbluelight0.0	&	1.5e-6	\\
eil51\_n19\_m18\_uncorr-similar-weights\_10	&	5573.695	&	5878.8	&	\cellbluelight0.0	&	0.0	&	10.5	&	\cellbluelight0.0	&	0.0	&	\cellbluelight0.0	&	0.0	\\
eil51\_n20\_m19\_multiple-strongly-corr\_01	&	518.189	&	4533.7	&	\cellbluelight0.6	&	0.6	&	11.1	&	14.1	&	1.4e-6	&	12.3	&	0.0	\\
eil51\_n20\_m19\_multiple-strongly-corr\_10	&	-	&	-	&	-	&	-	&	12.1	&	-	&	-	&	-	&	-	\\
eil51\_n20\_m19\_uncorr\_01	&	2092.673	&	2456.9	&	\cellbluelight0.0	&	0.0	&	8.7	&	\cellbluelight0.0	&	0.0	&	\cellbluelight0.0	&	0.0	\\
eil51\_n20\_m19\_uncorr\_10	&	3044.391	&	12776.0	&	\cellbluelight0.0	&	0.0	&	9.8	&	\cellbluelight0.0	&	0.0	&	\cellbluelight0.0	&	0.0	\\
eil51\_n20\_m19\_uncorr-similar-weights\_01	&	451.052	&	1007.7	&	\cellbluelight0.0	&	0.0	&	7.9	&	\cellbluelight0.0	&	0.0	&	\cellbluelight0.0	&	0.0	\\
eil51\_n20\_m19\_uncorr-similar-weights\_10	&	4169.799	&	15075.7	&	\cellbluelight0.0	&	0.0	&	9.4	&	\cellbluelight0.0	&	0.0	&	\cellbluelight0.0	&	0.0	\\
\hline
\end{tabular}
\end{adjustwidth}
\caption{Comparison between DP and the approximate approaches running in 10 minutes limits. Each approximate algorithm runs 10 times for each instance and use the average as the objective $Obj$. Gap is measured by $\frac{OPT - Obj}{OPT}\%$ and runtime (RT) is in second. The results of C5 and DP-S5 are obtained when they reach the time limit of 10 minutes per instance. Highlighted in blue are the best approximate results. DP runs out of memory for the instances without results.}
\label{tab:results2}
\end{table}

\section{Conclusion}\label{sec:conclusion}
The traveling thief problem (TTP) has attracted significant attention in recent years within the evolutionary computation community. In this paper, we have presented and evaluated exact approaches for the TTP based on dynamic programming, branch and bound, and constraint programming. We have used the exact solutions provided by our DP approach to evaluate the performance of current state-of-the-art TTP solvers. Our investigations show that they are obtaining in most cases (close to) optimal solutions. However, for a small fraction of tested instances we obverse a gap to the optimal solution of more than 10\%.

\section*{Acknowledgements}

This work was supported by the Australian Research councils through grants DP130104395 and DE160100850, and by the supercomputing resources provided by the Phoenix HPC service at the University of Adelaide.

\bibliographystyle{abbrvnat}
\bibliography{ttpdp}  

\end{document}